\documentclass[a4paper,sort&compress]{elsarticle}

\usepackage{
epsfig,psfrag,amsmath,amssymb,
scalefnt,wasysym}
\usepackage{color}
\usepackage{makeidx}
\usepackage{multirow}
\usepackage{longtable}
\usepackage{hyperref}

\definecolor{Black}{RGB}{0,0,0}
\definecolor{SkyBlue}{RGB}{200,225,255}

\newcommand{\GeV}{\mathrm{GeV}}

\newcommand{\gsim}{\;\rlap{\lower 3.5 pt \hbox{$\mathchar \sim$}} \raise 1pt
 \hbox {$>$}\;}
\newcommand{\lsim}{\;\rlap{\lower 3.5 pt \hbox{$\mathchar \sim$}} \raise 1pt
 \hbox {$<$}\;}

\newcommand*\circled[2]{\scalebox{#1}{\unitlength1ex\begin{picture}(2.5,2.5)
\put(0.75,0.75){\circle{1.6}}\put(0.7,0.7){\makebox(0,0){\tiny{#2}}}\end{picture}}}

\newcommand*\refm[1]{{\small (m\ref{#1})}}
 
\newcommand{\Mathematica}{{\tt Mathematica}}
\newcommand{\SLAM}{{\tt SLAM}}
\newcommand{\spheno}{{\tt SPheno}}
\newcommand{\softsusy}{{\tt SOFTSUSY}}
\newcommand{\suseflav}{{\tt SuSeFLAV}}
\newcommand{\suspect}{{\tt Suspect}}
\newcommand{\SQL}{{\tt SQL}}
\newcommand{\bash}{{\tt bash}}

\newcommand{\figpath}{figs/}

\newcounter{CounterMathematicaBox}
\newcommand{\MathematicaBox}[1]{\par\noindent\refstepcounter{CounterMathematicaBox}\\[5pt]\fcolorbox{Black}{SkyBlue}{\begin{minipage}[t]{0.90\linewidth}\centering\includegraphics[scale=0.8]{#1}\end{minipage}\begin{minipage}[b]{0.05\linewidth}\tiny(m\arabic{CounterMathematicaBox})\end{minipage}}\\[5pt]}
\newcommand{\MathematicaMiniBox}[1]{\par\noindent\refstepcounter{CounterMathematicaBox}\\[5pt]\fcolorbox{Black}{SkyBlue}{\begin{minipage}[t]{0.90\linewidth}\centering\includegraphics[scale=0.53]{#1}\end{minipage}\begin{minipage}[b]{0.05\linewidth}\tiny(m\arabic{CounterMathematicaBox})\end{minipage}}\\[5pt]}

\newenvironment{cautionbox}
{\par\noindent\begin{minipage}[c]{0.15\linewidth}\includegraphics[scale=1.0]{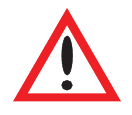}\end{minipage}\begin{minipage}[c]{0.85\linewidth}}
{\end{minipage}}

\newenvironment{smallercautionbox}
{\par\noindent\begin{minipage}[c]{0.15\linewidth}\includegraphics[scale=1.0]{\figpath caution.pdf}\end{minipage}\begin{minipage}[c]{0.83\linewidth}}
{\end{minipage}}

\allowdisplaybreaks

\makeindex

\begin{document}
\frontmatter

\title{\SLAM,\\ 
  a \Mathematica{} interface for\\
  SUSY spectrum generators
}

\author[Zeuthen]{Peter Marquard}
\ead{peter.marquard@desy.de}
\author[nz]{Nikolai Zerf}
\ead{zerf@ualberta.ca}

\address[Zeuthen]{Deutsches Elektronen Synchrotron DESY, Platanenallee 6, D15738 Zeuthen, Germany}
\address[nz]{Department of Physics, 
 University of Alberta,
  Edmonton AB T6G 2J1, Canada}
\date{}
\begin{abstract}
  We present and publish a \Mathematica{} package, which can be used to
  automatically obtain any numerical MSSM input parameter from SUSY
  spectrum generators, which follow the SLHA standard, like \spheno{},
  \softsusy{}, \suseflav{} or \suspect.  The package enables a very comfortable way
  of numerical evaluations within the MSSM using \Mathematica.  It
  implements easy to use predefined high scale and low scale scenarios
  like mSUGRA or $m_h^{\rm max}$ and if needed enables the user to
  directly specify the input required by the spectrum generators.  In
  addition it supports an automatic saving and loading of SUSY spectra
  to and from a SQL data base, avoiding the rerun of a spectrum
  generator for a known spectrum.  
\end{abstract}

\maketitle

\newpage

\section*{Program summary}

\begin{itemize}

\item[]{\it Title of program:}
  {\tt \SLAM}

\item[]{\it Available from:}\\
  {\tt
  http://www.ttp.kit.edu/Progdata/ttp13/ttp13-024/
  }

\item[]{\it Computer for which the program is designed and others on which it
    is operable:}
  Any computer where \Mathematica{} version~6 or higher is running providing \bash{} and {\tt sed}.

\item[]{\it Operating system or monitor under which the program has been
    tested:}
  Linux

\item[]{\it No. of bytes in distributed program including test data etc.:}
   $37\,060$ bytes.

\item[]{\it Distribution format:}
  source code

\item[]{\it Keywords:}
  SUSY, MSSM, SLHA, SQL, database, Mathematica, particle spectrum generation.

\item[]{\it Nature of physical problem:}
  Interfacing published spectrum generators for automated creation, saving and loading of SUSY particle spectra.

\item[]{\it Method of solution:}
  \SLAM{} automatically writes/reads SLHA spectrum generator input/output and is able to save/load generated data in/from a data base.

\item[]{\it Restrictions:}
  No general restrictions, specific restrictions are given in the manuscript.

\item[]{\it Typical running time:}
  A single spectrum calculation takes much less than one second on a modern PC.

\end{itemize}



\section{Introduction}
Although there has been no experimental evidence for the realization
of Supersymmetry (SUSY) yet, the Minimal Supersymmetric Standard
Model (MSSM) became very popular among particle physicists.  Many
calculations of elementary particle processes include the effects of
SUSY particles.  This applies to cosmological predictions like relic
density of dark matter from early universe, predictions of cross
sections for the direct production of SUSY particles at colliders like
the LHC and ILC, the calculation of radiative corrections due to
the presence of SUSY particles for Standard Model (SM) processes, for
example in flavour physics, and many more.

Compared to the SM a supersymmetric model is more predictive
because supersymmetry imposes many relations between different
parameters.  However, due to the fact that we do not live in a
supersymmetric world SUSY must be softly broken in a realistic  model
describing our world. Since the breaking mechanism is still unknown
many new unknown parameters arise in the broken model.


Once one makes an assumption for those unknown parameters or assumes a
certain breaking mechanism of SUSY, many parameters can be determined
directly from the knowledge of the measured SM parameters.  
For example, the Higgs mass is an independent parameter in the
SM.  In the MSSM this is no longer true, because after assuming a certain
SUSY breaking model it can be calculated from the knowledge of a few
parameters.  On the one hand this is a nice feature, but on the other
hand it leads to the problem of a consistent determination of all
relevant parameters like masses, mixing angles, and couplings in such a
model.  A collection of such a set of parameters will be called
spectrum in the following.

Fortunately this problem has already been solved and spectra can be
calculated auto\-matically using spectrum generators like
\spheno{}~\cite{Porod:2003um}, \softsusy{}~\cite{Allanach:2001kg}, \suseflav{}~\cite{Chowdhury:2011zr} or
\suspect{}~\cite{Djouadi:2002ze}.

In order to manage the identification of particle parameters and
couplings as well as option settings for programs in a common way, the
Supersymmetry Les Houches Accord (SLHA) has been
proposed~\cite{Skands:2003cj,Allanach:2008qq} and is used by many
programs including the mentioned ones\footnote{A list of programs
  using SLHA can be found in Ref.~\cite{SLHAWeb}.}.  Within the SLHA
the generation of a consistent spectrum can be specified by providing a
single input file ({\tt LesHouches.in}).  Providing this file leads,
after a run of the generator, to an output file ({\tt
  LesHouches.out}) containing the complete spectrum in the SLHA
notation.

If one is interested in only a few parameters of a single
spectrum one can easily extract them via copy and paste by hand.
However, this procedure is certainly not feasible if one needs to
extract many parameters of more than one spectrum for further
numerical evaluation.  In fact this is the situation one is
facing immediately when trying to create a plot in dependence of
high or low scale scenario parameters like $\tan{\beta}$.  That means
in order to be able to use the output of spectrum generators or any
other program using SLHA, one needs an interface providing an
automatic extraction of the relevant parameters from the output file
{\tt LesHouches.out}.

There already exist public interfaces written for {\tt
  C++}~\cite{1c++Web,2c++Web,LHPC,Belanger:2010st}, {\tt
  FORTRAN}~\cite{Hahn:2004bc,Hahn:2006nq,Belanger:2010st}, and {\tt
  Python}~\cite{pyslhaWeb}.

Due to its vast amount of implemented functions and with
increasing computer power and increasing computation speed,
\Mathematica{} became an attractive alternative for numerical evaluation
of analytic expressions obtained in SUSY models.  However, to the
authors' knowledge there is up to now no public implementation of an
interface to spectrum generators, automatically writing and reading
SLHA files with the goal to provide simple data sets
containing all needed parameters in \Mathematica{} for general\footnote{The \Mathematica{} package \verb~H3.m~~\cite{Kant:2010tf} comes with such an interface built-in
but it enables the automatic extraction of some special parameters needed in the package, only.} purpose.

This task is accomplished by \SLAM{} ({\bf S}upersymmetry {\bf L}es
Houches {\bf A}ccord with { \tt {\bf M}athematica}) 
which will be presented in detail throughout this publication.
A preliminary version of this package with name {\tt LHSQLDB} was already used in Refs.~\cite{Pak:2012xr} and~\cite{Kurz:2012ff}. 

This paper is organized as follows: for the impatient reader we
present in Section~2 typical usage examples which should
give a compact overview of what \SLAM{}  is capable to do.  In Section~3
we give detailed information about how to install and configure
the package properly before we give a full usage instruction in
Section~4.  Section~5 is dedicated to the internal structure of
\SLAM{}  and may help interested users to go beyond the ``black box''
model of this package.

\section{Teaser Examples}
To demonstrate, how the package should be used, we present some of the
implemented scenarios in this section. The full list of predefined
scenarios can be found in Tab.~\ref{TAB:PreimplementedModels}.
\subsection{Predefined $m_h^{\text{max}}$-scenario}\label{CHAP:PredefinedMhMaxScenario}
Once a spectrum generator is installed on the system and \SLAM{}  is
configured properly it is very easy to obtain a SLHA spectrum within
\Mathematica.  Install the package into the \Mathematica{} kernel by
loading the main file:

\MathematicaBox{\figpath loading.pdf}

Afterwards, one can generate spectra, e.g. in the predefined
$m_h^{\text{max}}$-scenario~\cite{Carena:2002qg} by using the central
function \verb|ObtainLesHouchesSpectrum|:

\MathematicaBox{\figpath mhmaxrun1.pdf}

With the displayed command we have requested a
$m_h^{\text{max}}$-spectrum using the spectrum generator \spheno.
Such a spectrum depends on $M_{A}$ and $\tan{\beta}$.  Because we did
not switch off the usage of a currently empty data base, \SLAM{}  
creates a new table in it.  Since there is no matching spectrum in the
data base the spectrum generator \spheno{} is run with the given parameters and
the spectrum stored in the data base.

Finally it returns a part of the spectrum in a list of replacement
rules.  
Since no parameters have been requested explicitly, 
 \SLAM{}  just returned a list of selected parameters, which
are defined in a default request list.

Assuming we are only interested in the mass of the lightest Higgs boson
$M_{h^0}$ within the same scenario we can use:

\MathematicaBox{\figpath mhmaxrun2.pdf}
\label{MB:MhMaxRun2}

where we have used the SLHA identification for the on-shell Higgs
mass, settled in the Block \verb|"MASS"| with key entry $25$.  The
choice of the symbol \verb|Mh0| is completely arbitrary, up to the
restriction that it has to be an undefined symbol.  
This allows the user to use his own notation in
\Mathematica.  The \verb|Real| statement tells \SLAM{}  that it should
treat the parameter during the internal processing as a real valued
number.  Note that \SLAM{}  did not rerun \spheno{}  to obtain the Higgs
mass, but just retrieved the value directly from the data base, where it has been stored during the
previous call of the function \verb|ObtainLesHouchesSpectrum|.

In order to see the default value of the option \verb|InputRequest|
one can use:

\MathematicaBox{\figpath defaultinputrequest.pdf}
\label{MB:defaultinputrequest}
 
With the given output it is straightforward to 
customize the  \verb|InputRequest| option values for the user's needs.

Beside the $m_h^{\text{max}}$ scenario the no-mixing, gluophobic and
small $\alpha_{\rm eff}$ scenarios, which were proposed together with
the $m_h^{\text{max}}$ scenario in Ref.~\cite{Carena:2002qg}, are
already implemented and have the same input variables $M_{A}$ and
$\tan{\beta}$.  They are used when setting the \verb|Model| option to
\verb|"nomix"|, \verb|"smallalpha"|, or \verb|"gluophob"|.

\subsection{Predefined mSUGRA-scenario}
The mSUGRA-scenario exists as a predefined model and can be called as
follows:

\MathematicaBox{\figpath mSUGRA.pdf}

The shown values for the five input parameters represent their default
values.  One should be aware that those values are already excluded experimentally 
and their allowed values are much higher.

\subsection{Predefined p19MSSM plane I/II scenario}
Since more and more space of the mSUGRA scenario has been excluded at
the LHC, there has been a proposal of many scenarios in
Ref.~\cite{AbdusSalam:2011fc} taking into account 
exclusions by this non-observation of SUSY particles.  If there is no
interest in a SUSY breaking mechanism motivated by some high scale
physics, one can just define all the SUSY breaking parameters at a low
scale.  With certain assumptions one can reduce the number of these
parameters to 19.

Two different planes have been defined in such a phenomenological 19
parameter MSSM~\cite{AbdusSalam:2011fc}:
\begin{itemize}
\item p19MSSM plane I (\verb|"p19MSSM1"|) in dependence of
  $M_1$(\verb|M1|) and $M_3$(\verb|M3|).
 \item p19MSSM plane II (\verb|"p19MSSM2"|) depends on
   $M_1$(\verb|M1|) and a common soft slepton mass breaking parameter
   $m_{\tilde{l}}$(\verb|mslepton|) of the first two generations.
\end{itemize}
Both scenarios are implemented in \SLAM.
We just show an example for the first one:

\MathematicaBox{\figpath p19MSSM1.pdf}

To provide a reasonable light Higgs mass, we show the \spheno{}  result
for the scenario point p19MSSM1.13.  The very first points in the
sequence (p19MSSM1.1, p19MSSM1.2, \dots) seem to be already excluded
if one assumes that the new particle discovered at the LHC is the
lightest Higgs boson.

\subsection{User-defined scenarios}\label{CHAP:SelfDefinedScenarios}
On the one hand predefined scenarios are quite useful, because the user
does not need to worry about all the required input parameters, on the
other hand they are very restrictive.  To remove this disadvantage,
\SLAM{}  can be used to generate spectra for any user-defined scenario
one can think of.

This is done by handing all relevant input parameters to \SLAM{}
directly via the option \verb|Model| of the function \verb|ObtainLesHouchesSpectrum|.  In the example
below we generate the same mSUGRA spectrum like two sections ago but
this time using direct model input for the underlying scenario:

\MathematicaBox{\figpath directmodel.pdf}

In the first line we defined our own model in variable \verb|MyModel|.
In this example, we use a
minimal definition of the model, only giving the definition of needed
entry values which are indicated by the \verb|EV| symbols. In
addition, it is possible to define comments for every entry, which
can improve the readability of the SLHA files.

As can be seen from the output, the requested spectrum was already
stored in the data base, which is the case since it corresponds to the
predefined mSUGRA scenario used before.

\begin{cautionbox}
  We have to add the warning that when using user-defined scenarios,
  the user has to make sure that the input provided is sufficient for
  the spectrum generators to run correctly.  \SLAM{}  will not check
  the provided input for its consistency. 
\end{cautionbox}


\section{Prerequisites \& Installation}
\subsection{Prerequisites}
To be able to use \SLAM{}  one needs:
\begin{itemize}
 \item \Mathematica{}  of version $6.0$ or higher.
 \item A UNIX operating system providing \bash{}  and {\tt sed}.
\end{itemize}
And at least one of the spectrum generators listed in the following table.\\

\begin{tabular}{ccc}
Generator   & Weblink & Version \\
\hline
\spheno{}   & {\small http://spheno.hepforge.org/}     & 3.2.0\\
\softsusy{} & {\small http://softsusy.hepforge.org/}  &  3.3.4\\
\suseflav{} & {\small https://suseflav.hepforge.org/} &  1.2.0\\
\multirow{2}{*}{\suspect{}}& {\small http://www.coulomb.univ-montp2.fr}  & \multirow{2}{*}{2.4.3}\\
                           & {\small /perso/jean-loic.kneur/Suspect/}    &                       \\
\hline\\
\end{tabular}

They can be obtained from the stated websites.
The versions indicated have been used for this publication\footnote{using different versions may result in slightly different output}, but SLAM is not restricted to them.

A running \SQL{}  server may be very convenient for quickly storing and
loading spectrum data.  But it is not mandatory, because
\Mathematica{} has a built-in data base server\footnote{which however slows down when dealing with a large number (>1000) of saved spectra}.
\subsection{Installation}
Installing \SLAM{}  is quite simple:
\begin{itemize}
 \item{ Download the package from~\cite{SLAM_DL_LINK}.}  
 \item{ Unpack the tarball and put the two files:
\begin{itemize}
 \item \verb|SLAM.m|
 \item \verb|SLAM.config.m|
\end{itemize}
in a common folder.  }
\item Open \verb|SLAM.config.m| with any text editor and adjust the
  given entries appropriately.
\end{itemize}
To clarify the last step, we give detailed information about every
setting contained in \verb|SLAM.config.m|, although the contained
comments should be sufficient to adjust the settings
without manual.

The configuration file \verb|SLAM.config.m| is -- like the
extension suggests -- obeying \Mathematica{} syntax and any change
in this file must not break it.  In the following we describe the
contained settings.
{\small
\LTpre=\fill
\begin{longtable}{|c|c|c|}
  \hline
  \multirow{2}{*}{Variable}& Symbol              & \multirow{2}{*}{Description}\\
                           & Default Value       &                             \\
  \hline
  \endfirsthead
  \hline
  \multirow{2}{*}{Variable}& Symbol              & \multirow{2}{*}{Description}\\
                           & Default Value       &                             \\
  \hline
  \endhead
  \hline
  \caption{\label{TAB:DefaultValues} Default definition of numerical values used in \SLAM{}  in proper powers of $\GeV$.
  The circled numbers show the ID of the corresponding model, where the specific default value is used. See Tab.~\ref{TAB:PreimplementedModels} for more details.}
  \endfoot
  \hline
  \caption[]{(continued)}
  \endlastfoot
  \multirow{2}{*}{\verb~invalphaValue~}& $\alpha^{-1,\overline{\mathrm MS}}(m_Z)$   & \multirow{2}{*}{inverse fine structure constant}                               \\*
                                       & $1.279340000\cdot10^2$             &                                                                                       \\
  \multirow{2}{*}{\verb~GFValue~}      & $G_F$                              & \multirow{2}{*}{Fermi-constant}                                                       \\*
                                       & $1.16637\cdot10^{-5}$              &                                                                                       \\
  \multirow{2}{*}{\verb~asmzValue~}    & $\alpha_s^{(5),\overline{\mathrm MS}}(m_Z)$& \multirow{2}{*}{QCD fine structure constant}                                  \\*
                                       & $0.1184$                           &                                                                                       \\
  \multirow{2}{*}{\verb~MwValue~}      & $m_W$                              & \multirow{2}{*}{W boson mass}                                                         \\*
                                       & $80.399$                           &                                                                                       \\
  \multirow{2}{*}{\verb~MzValue~}      & $m_Z$                              & \multirow{2}{*}{Z boson mass}                                                         \\*
                                       & $91.200$                           &                                                                                       \\
  \multirow{2}{*}{\verb~MbMSbarValue~} & $m_b^{\overline{\mathrm MS}}(m_b)$ & \multirow{2}{*}{b quark mass}                                                         \\*
                                       & $4.25$                             &                                                                                       \\
  \multirow{2}{*}{\verb~MtOSValue~}    & $M_t$                              & \multirow{2}{*}{t quark pole mass}                                                    \\*
                                       & $173.3$                            &                                                                                       \\ 
  \multirow{2}{*}{\verb~MtauValue~}    & $m_{\tau}$                         & \multirow{2}{*}{$\tau$ lepton pole mass}                                              \\*
                                       & $1.777$                            &                                                                                       \\
  \multirow{2}{*}{\verb~m0Value~}      & $m_0$                              & \multirow{2}{*}{common scalar mass \circled{1.1}{1}\circled{1.1}{3}}                  \\*
                                       & $1.0\cdot10^2$                     &                                                                                       \\
  \multirow{2}{*}{\verb~m12Value~}     & $m_{1/2}$                          & \multirow{2}{*}{common gaugino mass \circled{1.1}{1}}                                 \\*
                                       & $2.500\cdot10^2$                   &                                                                                       \\
  \multirow{2}{*}{\verb~signmuValue~}  & $\text{sgn}(\mu)$                  & \multirow{2}{*}{sign of bilinear $H_1H_2$ coupling \circled{1.1}{1}-\circled{1.1}{3}} \\*
                                       & $1.000$                            &                                                                                       \\*
  \multirow{2}{*}{\verb~A0Value~}      & $A$                                & \multirow{2}{*}{unified trilinear coupling \circled{1.1}{1}}                          \\*
                                       & $-1.000\cdot10^2$                  &                                                                                       \\
  \multirow{2}{*}{\verb~lambdaValue~}  & $\Lambda$                          & \multirow{2}{*}{scale of soft SUSY breaking \circled{1.1}{2}}                         \\*
                                       & $40.\cdot10^3$                     &                                                                                       \\
  \multirow{2}{*}{\verb~mmessValue~}   & $M_{\rm mess}$                     & \multirow{2}{*}{overall messenger scale \circled{1.1}{2}}                             \\*
                                       & $80.\cdot10^3$                     &                                                                                       \\
  \multirow{2}{*}{\verb~nmessValue~}   & $N_5$                              & \multirow{2}{*}{messenger index \circled{1.1}{2}}                                     \\*
                                       & $3.$                               &                                                                                       \\
  \multirow{2}{*}{\verb~m32Value~}     & $m_{3/2}$                          & \multirow{2}{*}{gravitino mass \circled{1.1}{3}}                                      \\*
                                       & $60.\cdot10^3$                     &                                                                                       \\
  \multirow{2}{*}{\verb~MAValue~}      & $m_A$                              & \multirow{2}{*}{$A^0$ Higgs boson mass \circled{1.1}{4}- \circled{1.1}{14}}           \\*
                                       & $200.$                             &                                                                                       \\
  \multirow{2}{*}{\verb~tanbetaValue~} & $\tan{\beta}$                      & \multirow{2}{*}{$v_2/v_1$ \circled{1.1}{1}- \circled{1.1}{15}}                        \\*
                                       & $10.$                              &                                                                                       \\
  \multirow{2}{*}{\verb~muValue~}      & $\mu$                              & \multirow{2}{*}{bilinear $H_1H_2$ coupling \circled{1.1}{15}}                         \\*
                                       & $1.5\cdot10^3$                     &                                                                                       \\
  \multirow{2}{*}{\verb~M1Value~}      & $M_1$                              & \multirow{2}{*}{gaugino soft breaking mass \circled{1.1}{16}- \circled{1.1}{17}}       \\*
                                       & $300.$                             &                                                                                       \\
  \multirow{2}{*}{\verb~M3Value~}      & $M_3$                              & \multirow{2}{*}{gaugino soft breaking mass \circled{1.1}{16}}                         \\*
                                       & $360.$                             &                                                                                       \\
  \multirow{2}{*}{\verb~msleptonValue~}& $m_{\tilde{l}}$                    & \multirow{2}{*}{common slepton breaking mass \circled{1.1}{17}}                       \\*
                                       & $400.$                             &                                                                                       \\
  \multirow{2}{*}{\verb~xBFS2013Value~}& $x$                                & \multirow{2}{*}{slope parameter $x$ \circled{1.1}{21}}                                \\*
                                       & $0.$                               &                                                                                       \\
  \multirow{2}{*}{\verb~yBFS2013Value~}& $y$                                & \multirow{2}{*}{slope parameter $y$ \circled{1.1}{21}}                                \\*
                                       & $0.$                               &                                                                                       \\
  \multirow{2}{*}{\verb~muRenValue~}   & $\mu_{\rm ren}$                    & \multirow{2}{*}{renormalization/output scale}                                         \\*
                                       & $173.3$                            &                                                                                       \\
\end{longtable}
}
The first set of options has to be adjusted to reflect the user's setup
\begin{itemize}
\item{\verb|ProgramPath["spheno"]="/.../SPheno-3.2.0/bin/SPheno"|\\
    \verb|ProgramPath["softsusy"]="/.../softsusy-3.3.4/bin/softpoint.x"|\\
    \verb|ProgramPath["suseflav"]="/.../SuSeFLAV-1.2/bin/suseflavslha"|\\
    \verb|ProgramPath["suspect"]="/.../suspect2/suspect2"|\\
    define the path to the \spheno{}, \softsusy{}, \suseflav{} and \suspect{} executables
    on the file system as a string.}
\item{\verb|SQLDataBaseTypeValue->"hsqldb"|\\
    sets the data base type to the default data base server included
    in \Mathematica.
    To connect to a dedicated \SQL{}  server this settings might change to the following:\\
    \verb|SQLDataBaseTypeValue->"MySQL(Connector/J)"|\\
    To be more specific this setting is used as an argument when connecting to the data base with the command:\\
    \verb|OpenSQLConnection[JDBC[SQLDataBaseTypeValue,...]...]|.\\
    One can use the \Mathematica{}  built-in connection tool\\
    \verb|Needs["DatabaseLink`"];|\\
    \verb|OpenSQLConnection[]|\\
    to find/set the proper definition of the needed values here and
    below.  More details can be found in the documentation centre of
    \Mathematica{}.  }\label{ITM:SQLDataBaseTypeValue}
\item{\verb|SQLDataBaseOptionsValue->|\\
     \verb|{"Name" -> "Spectrum", "Username" ->Environment["USER"]}|\\
    contains a list of options relevant for the data base.\\
    When using a \SQL{}  server the list might look as follows:\\
    \verb|{"Username" -> Environment["USER"], "Password" -> "xxx",|\\
    \verb|"Catalog" -> "..."}|\\
    This list is used when connecting to the data
    base: \verb|OpenSQLConnection[JDBC[...],SQLDataBaseOptionsValue]|
  }\label{ITM:SQLDataBaseOptionsValue}
\item{\verb|SQLDataBaseValue ->|\\
     \verb|(Environment["HOME"]<>"/.SLAM/database/db")|\\
    gives the link to the data base as a string.  When using the built-in \Mathematica{} data base, this is the location on the system
    where \Mathematica{} will store its data base files.
    In case a dedicated \SQL{}  server is used, this string contains its address including port information:\\
    \verb|SQLDataBaseValue -> "sqlserver.someaddress.de:3306"|\\
    The option is used as a second argument during the connection to the data base:\\
    \verb|OpenSQLConnection[JDBC[...,SQLDataBaseValue],...]|
  }\label{ITM:SQLDataBaseValue}
\end{itemize}
The next set of options can be left at their default values, but can be
changed if needed
\begin{itemize}
\item{\verb|SQLVarCharLengthValue -> 50|\\
 gives the maximum length of strings saved in the data base in the \verb|"VARCHAR"| format.}\label{ITM:SQLVarCharLengthValue}
 \item{\verb|InputFilePathValue->|\\
       \verb|(Environment["HOME"]<>"/.SLAM/tmp/LesHouches.in")|\\
 contains the name and location of temporary SLHA input files as a string. 
 }\label{ITM:InputFilePathValue}
 \item{\verb|OutputFilePathValue->|\\
       \verb|(Environment["HOME"]<>"/.SLAM/tmp/LesHouches.out")|\\
contains the name and location of temporary SLHA output files as a string. 
}\label{ITM:OutputFilePathValue}
\item{
The default values of the parameters are defined. All parameters and
their default values are listed together with a brief
description in
Tab.~\ref{TAB:DefaultValues}.
}
\item{After the default numerical values are set, one is able
    to configure the default input from \SLAM{}  for further
    processing.  For example:
    \verb|LesHouchesInputRequest = {"MASS"->{25 ->{Mh0, Real}}};|\\
    will just return the lightest Higgs boson mass by default.  A
    more general example for the list on the right side of the
    definition can be taken from Mbx.~\refm{MB:defaultinputrequest}.
  }
\item{\verb|UseMinimalFlavourViolation=True;|\\
    Sets the used set of declarations done in the bottom of the configuration file
    to the minimal flavour violation one.  If one works within minimal flavour
    violating models one should keep this setting.  All the predefined
    scenarios work within this framework.  If one works beyond minimal
    flavour violation, one may set the variable to \verb|False|.  }
\end{itemize}
The following settings should only be changed by experienced users,
who intend to extend \SLAM{}.
\begin{itemize}
\item{\verb|LesHouchesInputSQLFormats= { ... }|\\
    Contains declarations of input parameters, which are required by
    spectrum generators.  These declarations are used to define the
    data table layout used in the \SQL{}  data base.  Possible data types are
    \verb|String|, \verb|Integer| and \verb|Real|.

A change of the declarations requires a reset of the used data base.
\SLAM{}  requires a proper set of declarations in order to be able to
use a particular data base table.  }
\item{\verb|LesHouchesOutputSQLFormats= { ... }|\\
    Contains declarations of output parameters, which are generated by
    spectrum generators.  These declarations are used to define the
    data table layout used in the \SQL{}  data base.  Possible data types are
    \verb|String|, \verb|Integer| and \verb|Real|.

A change of the declarations requires a reset of the used data base.
\SLAM{}  requires a proper set of declarations
in order to be able to use a particular data base.
}
\end{itemize}

If one wants to load \SLAM{}  in a more convenient way with the command
\verb|Needs["SLAM`"];|, the path to the folder where
\verb|SLAM.m| is located can be added to the list of folder names stored in
the global \verb|$Path| variable.  This can be done for example in the
\verb|~/.Mathematica/Kernel/init.m| file by the following command:
\begin{verbatim}
$Path=Join[$Path,{"/some/path/SLAM"}];
\end{verbatim}

Concerning the installation and configuration of a \SQL{}  server, we
recommend to follow the instructions given in the particular server
manual.  However, we want to add some comments about how to speed up
the \SQL{}  server in the following.

The access to the data base can be greatly improved by adding a key or an
index to the most important columns of the table. 
For example in the case of the $m_h^{\rm max}$
scenario the most obvious choice would be to at least add a key to
the columns containing the  input parameters $m_A$ and $\tan{\beta}$. 
For other scenarios a different choice might be more suitable\footnote{See Tab.~\ref{TAB:PreimplementedModels} for examples of free scenario parameters.}. 
Keys can be defined directly after the initial declaration of the
layout of the data base or added at any later time at almost no
cost. 
To define a key using \Mathematica{}  considering for example the $m_h^{\rm max}$ scenario one can use the command 
\begin{verbatim}
SQLExecute[$LesHouchesSQLConnectionList[[1]], #] & /@ 
{"alter table SPECTRATABLE add key (I_EXTPAR_26);" (*MA*), 
 "alter table SPECTRATABLE add key (I_MINPAR_3);" (*tanbeta*)};
\end{verbatim}
which requires an active data base connection stored in {\tt \$Les\-Houches\-SQL\-ConnectionList}.
An active data base connection can be obtained best by calling {\tt Obtain\-Les\-Houches\-Spectrum[]} with the option {\tt Keep\-SQL\-Connection} set to {\tt True} once,
before executing the command above.
Since the \Mathematica{}  function {\tt SQL\-Execute[]} used in this
example simply executes the given \SQL{}  command, one can 
easily read off the necessary \SQL{}  commands enclosed as strings.
If one is interested in the inverse problem, i.e. looking up
spectra which e.g. have a Higgs boson in a certain mass range, it is
helpful to define a key for the respective column.
For a definition of column names in the data base table see Section~\ref{CHAP:DataBaseTableLayout} on page~\pageref{CHAP:DataBaseTableLayout}.

\section{Usage Instructions}
In this section we try to give detailed instructions about how to
use the \SLAM{}  package.  Mainly this comes down to the knowledge how
to use the function {\tt Obtain\-Les\-Houches\-Spectrum}.  But also
the built-in function {\tt Read\-LesHouches\-Spectrum\-File} might be
of general interest because it can automatically import any SLHA file
into \Mathematica.  The same holds for the function {\tt
  Write\-LesHouches\-File}, which automatically writes SLHA files
according to the input provided in \Mathematica.

\subsection{Using {\tt Obtain\-Les\-Houches\-Spectrum}}
\index{Functions!{\tt Obtain\-Les\-Houches\-Spectrum}} The function
{\tt Obtain\-Les\-Houches\-Spectrum} is completely controlled via options.  One might even call the function without giving an
explicit option:

\MathematicaBox{\figpath nooptionsrun.pdf}
 
This calls {\tt Obtain\-Les\-Houches\-Spectrum} with all options
set to their default values. A list of all options can be obtained
using:

\MathematicaMiniBox{\figpath alloptions.pdf}
\label{MB:AllOptions}
 
One can get further information about a single
option by the question mark operation as follows:

\MathematicaBox{\figpath questionmarkoperation.pdf}

The default option value can be checked with:

\MathematicaBox{\figpath defaultoptionasmz.pdf}
 
and set via:

\MathematicaBox{\figpath settingdefaultoption.pdf}

In principle one can just go through all options using the given
commands for each option in order to get to know all of them.

Since this may be tedious, 
we just pick out classes of options and discuss them step by step.

The first class of options is part of predefined scenarios.  In
detail they are just the free parameters left in the corresponding
scenario and listed in Tab.~\ref{TAB:PreimplementedModels}.
\begin{table}[t]\small\centering
\begin{tabular}{|c|c|c|c|c|}\hline
  Model & Option value & ID/Ref.& Free Parameters\\
  \hline
\multirow{2}{*} {  mSUGRA\index{Model!mSUGRA} }                                            & \multirow{2}{*} { \verb~"msugra"~  }           & \multirow{2}{*} {\circled{1.1}{1}} & \verb&m0&\index{Options!\verb&m0&}, \verb&m12&\index{Options!\verb&m12&}, \verb&A0&\index{Options!\verb&A0&}, \\
&&&\verb&signmu&\index{Options!\verb&signmu&}, \verb&tanbeta&\index{Options!\verb&tanbeta&}\\
\multirow{2}{*} {  mGMSB\index{Model!mGMSB}}                                               & \multirow{2}{*} {\verb~"mgmsb"~ }             & \multirow{2}{*} {\circled{1.1}{2}}& \verb&lambda&\index{Options!\verb&lambda&}, \verb&mmess&\index{Options!\verb&mmess&}, \\
&&&\verb|tanbeta|, \verb|signmu|, \verb&nmess&\index{Options!\verb&nmess&}\\
  mAMSB\index{Model!mAMSB}                                                & \verb|"mamsb"|             & \circled{1.1}{3}& \verb|m0|, \verb&m32&\index{Options!\verb&m32&}, \verb|tanbeta|, \verb|signmu|\\
  $m_h^{\rm max}$\index{Model!$m_h^{\rm max}$}                           & \verb|"mhmax"|             & \circled{1.1}{4}~\cite{Carena:2002qg}& \verb&MA&\index{Options!\verb&MA&}, \verb|tanbeta|\\
  no-mixing\index{Model!no-mixing}                                       & \verb|"nomix"|             & \circled{1.1}{5}~\cite{Carena:2002qg}& \verb|MA|, \verb|tanbeta|\\
  gluophobic\index{Model!gluophobic}                                     & \verb|"gluophob"|          & \circled{1.1}{6}~\cite{Carena:2002qg}& \verb|MA|, \verb|tanbeta|\\
  small $\alpha_{\rm eff}$\index{Model!small $\alpha_{\rm eff}$}         & \verb|"smallalpha"|        & \circled{1.1}{7}~\cite{Carena:2002qg}& \verb|MA|, \verb|tanbeta|\\
  updated $m_h^{\rm max}$\index{Model!updated $m_h^{\rm max}$}           & \verb|"mhmaxup"|           & \circled{1.1}{8}~\cite{Carena:2013qia}& \verb&MA&\index{Options!\verb&MA&}, \verb|tanbeta|\\
  $m_h^{\rm mod+}$\index{Model!$m_h^{\rm mod+}$}                         & \verb|"mhmod+"|            & \circled{1.1}{9}~\cite{Carena:2013qia}& \verb&MA&\index{Options!\verb&MA&}, \verb|tanbeta|\\
  $m_h^{\rm mod-}$\index{Model!$m_h^{\rm mod-}$}                         & \verb|"mhmod-"|            & \circled{1.1}{10}~\cite{Carena:2013qia}& \verb&MA&\index{Options!\verb&MA&}, \verb|tanbeta|\\
  light stop\index{Model!light stop}                                   & \verb|"lightstop"|         & \circled{1.1}{11}~\cite{Carena:2013qia}& \verb&MA&\index{Options!\verb&MA&}, \verb|tanbeta|\\
  light stau\index{Model!light stau}                                    & \verb|"lightstau"|         & \circled{1.1}{12}~\cite{Carena:2013qia}& \verb&MA&\index{Options!\verb&MA&}, \verb|tanbeta|\\
  light stau ($\Delta_{\tau}$)\index{Model! light stau ($\Delta_{\tau}$)} & \verb|"lightstaudeltatau"| & \circled{1.1}{13}~\cite{Carena:2013qia}& \verb&MA&\index{Options!\verb&MA&}, \verb|tanbeta|\\
  $\tau$-phobic\index{Model!$\tau$-phobic}                               & \verb|"tauphobic"|         & \circled{1.1}{14}~\cite{Carena:2013qia}& \verb&MA&\index{Options!\verb&MA&}, \verb|tanbeta|\\
  low-$m_h$\index{Model!low-$m_h$}                                       & \verb|"lowmh"|             & \circled{1.1}{15}~\cite{Carena:2013qia}& \verb&mu&\index{Options!\verb&mu&}, \verb|tanbeta|\\
  p19MSSM I\index{Model!p19MSSM I}                                      & \verb|"p19MSSM1"|          & \circled{1.1}{16}~\cite{AbdusSalam:2011fc}& \verb&M1&\index{Options!\verb&M1&}, \verb&M3&\index{Options!\verb&M3&}\\
  p19MSSM II\index{Model!p19MSSM II}                                    & \verb|"p19MSSM2"|          & \circled{1.1}{17}~\cite{AbdusSalam:2011fc}& \verb|M1|, \verb&mslepton&\index{Options!\verb&mslepton&}\\
  374345\index{Model!374345}                                            & \verb|"374345"|            & \circled{1.1}{18}~\cite{Cahill-Rowley:2013gca}&  ~ \\
  401479\index{Model!401479}                                            & \verb|"401479"|            & \circled{1.1}{19}~\cite{Cahill-Rowley:2013gca}&  ~ \\
  1046838\index{Model!1046838}                                            & \verb|"1046838"|            & \circled{1.1}{20}~\cite{Cahill-Rowley:2013gca}&  ~ \\
  2342344\index{Model!2342344}                                            & \verb|"2342344"|            & \circled{1.1}{21}~\cite{Cahill-Rowley:2013gca}&  \verb&xBFS2013&\index{Options!\verb&xBFS2013&}, \verb&yBFS2013&\index{Options!\verb&yBFS2013&} \\
  2387564\index{Model!2387564}                                            & \verb|"2387564"|            & \circled{1.1}{22}~\cite{Cahill-Rowley:2013gca}&  ~ \\
  2750334\index{Model!2750334}                                            & \verb|"2750334"|            & \circled{1.1}{23}~\cite{Cahill-Rowley:2013gca}&  ~ \\
  \hline
\end{tabular}
\caption{\label{TAB:PreimplementedModels}List of all predefined models available in \SLAM.
  Use the strings listed in the Option value column as option value for the {\tt Model} in order to select the corresponding model.
  All free parameters of each model are shown in the last column.}
\end{table}
Removing these options from the list in Mbx.~\refm{MB:AllOptions} the
number of unknown options reduces by 16.

The next class of options is built up by the SM parameters
which need to be provided to the spectrum generators. 
They are listed in Tab.~\ref{TAB:SMParameterOptions}

{\small
\begin{longtable}{|c|c|c|}
  \hline
  \multirow{2}{*}{Variable}& Symbol              & \multirow{2}{*}{Description}\\
                           & Default Value       &                             \\
  \hline
  \endfirsthead
  \hline
  \multirow{2}{*}{Variable}& Symbol              & \multirow{2}{*}{Description}\\
                           & Default Value       &                             \\
  \hline
  \endhead
  \hline
  \caption{(continued)}
  \endfoot
  \hline
  \caption[]{\label{TAB:SMParameterOptions}List of options used to specify SM parameters needed by the spectrum generators. All values are given in proper powers of $\GeV$.}
  \endlastfoot
  \multirow{2}{*}{\verb~invalpha~}& $1/\alpha^{\overline{MS}}(m_Z)$\index{Options!\verb&invalpha&} & \multirow{2}{*}{inverse fine structure constant}  \\
                                  & $1.279340000\cdot10^2$                                          &                                                      \\*
  \multirow{2}{*}{\verb~GF~}      & $G_F$\index{Options!\verb&GF&}                                  & \multirow{2}{*}{Fermi-constant}                      \\
                                  & $1.16637\cdot10^{-5}$                                           &                                                      \\*
  \multirow{2}{*}{\verb~asmz~}    & $\alpha_s^{(5),\overline{MS}}(m_Z)$\index{Options!\verb&asmz&}  & \multirow{2}{*}{QCD fine structure constant}         \\
                                  & $0.1184$                                                        &                                                      \\*
  \multirow{2}{*}{\verb~Mw~}      & $m_W$\index{Options!\verb&Mw&}                                  & \multirow{2}{*}{W boson mass}                        \\
                                  & $80.399$                                                        &                                                      \\*
  \multirow{2}{*}{\verb~Mz~}      & $m_Z$\index{Options!\verb&Mz&}                                  & \multirow{2}{*}{Z boson mass}                        \\
                                  & $91.200$                                                        &                                                      \\*
  \multirow{2}{*}{\verb~mb~}      & $m_b^{\overline{MS}}(m_b)$\index{Options!\verb&mb&}             & \multirow{2}{*}{b quark mass}                        \\
                                  & $4.25$                                                          &                                                      \\*
  \multirow{2}{*}{\verb~MtOS~}    & $M_t$\index{Options!\verb&MtOS&}                                & \multirow{2}{*}{t quark pole mass}                   \\
                                  & $173.3$                                                         &                                                      \\* 
  \multirow{2}{*}{\verb~Mtau~}    & $m_{\tau}$\index{Options!\verb&Mtau&}                           & \multirow{2}{*}{$\tau$ lepton pole mass}              \\
                                  & $1.777$                                                         &                                                      \\*
  \multirow{2}{*}{\verb~muRen~}   & $\mu_{\rm ren}$\index{Options!\verb&muRen&}                     & \multirow{2}{*}{renormalization/output scale}        \\
                                  & $173.3$                                                         &                                                      \\*
\end{longtable}
}
This further reduces the amount of unknown options by 9.

The next class is built up by options used to steer the processing.
\begin{table}[t]
\small\centering
\begin{tabular}{|c|c|c|}
\hline
Option & Default Value \\
\hline
\verb|InputFilePath| & see \verb|InputFilePathValue| p\pageref{ITM:InputFilePathValue}\\
\verb|InputRequest|& see Mbx.~\refm{MB:defaultinputrequest} on p\pageref{MB:defaultinputrequest}\\
\verb|Model|& \verb|"msugra"|\\
\verb|OutputFilePath|& see \verb|OutputFilePathValue| p\pageref{ITM:OutputFilePathValue}\\
\verb|RemoveTemporaryFiles|& \verb|True|\\
\verb|Silent|& \verb|False|\\
\verb|SpectrumGenerator|& \verb|"spheno"| \\
\verb|WorkingFolder|& \verb|Automatic| $\rightarrow$ folder of \verb|InputFilePath| \\
\hline
\end{tabular}
\caption{\label{TAB:ControlOptions}List of options used to control the core operation of \SLAM.}
\end{table}
A list of these options including their default values are given in Tab.~\ref{TAB:ControlOptions}.
In the following we give more detailed information on each of these options:
\begin{itemize}
 \item{\verb&InputFilePath&\index{Options!\verb&InputFilePath&}/\verb&OutputFilePath&\index{Options!\verb&OutputFilePath&}\\
 The path including the name of SLHA input/output file as a string.
 \begin{cautionbox}
If more than one \SLAM{} sessions are running at the same time using  a shared file system, 
one should include the process ID in those paths, in order to avoid different processes writing and reading from the same file.
\end{cautionbox}}
\item{\verb&InputRequest&\index{Options!\verb&InputRequest&}~\label{ITM:InputRequest}\\
Defines the data, which should be returned:
\begin{itemize}
\item{ \verb|All|\\
    returns everything that is available in an internal SLHA format,
    built up from lists and replacement rules.  For more details about
    the used format see
    Section~\ref{CHAP:InternalRepresentationfortheSLHA}.

In fact the output still depends on the source of the data:

If \SLAM{}  got the data from a SLHA-file written by a generator, the
entry values are still kept as strings, even though they are integers or
real numbers, because \SLAM{} does not know, which format
transformation should be applied.

If \SLAM{}  loaded the data from the data base,
all entry values will have their proper format.
}
\item{\verb|AllFormated|\\
    Returns only blocks of entry values, which were declared in {\tt
      SLAM.config.m}.  Moreover \SLAM{}  tries to convert all data
    to the proper format contained in the declaration.  }
\item{{\sc Direct input}\\
    Can be used to handle direct input, which selects
    certain data to be returned as a list of replacement rules, where
    the symbol of the left hand side of each rule can be freely
    chosen.  Examples for such selections can be found in
    Section~\ref{CHAP:PredefinedMhMaxScenario} in
    Mbx.~\refm{MB:MhMaxRun2}.  The default value for this option is
    printed in Mbx.~\refm{MB:defaultinputrequest}, where only
    entry values (\verb|EV|) are selected.  It is also possible to
    select block comments (\verb|BC|), entry value comments
    (\verb|EC|) and block information (\verb|BI|) as follows:
\begin{verbatim}
{"MASS" ->{BC       -> {MASSComment, String},
           {25, EC} -> {Mh0Comment, String}},
 "MSOFT"->{BI       -> {muRen, Real}}};
\end{verbatim}
    Note that if  data is requested, that was not declared in the data
    base table, \SLAM{}  will print a warning and abort the process,
    if the data base is in use.  If this is not the case, the data
    will be returned as {\tt \$Missing\-Data} in case the data is missing.  }
\end{itemize}
}
\item{ \verb&WorkingFolder&\index{Options!\verb&WorkingFolder&}~\\
    determines the place where temporary files for and from spectrum
    generators are created.  If one keeps the default option value,
    given by \verb|Automatic|, \SLAM{}  will extract the directory
    from the \verb|InputFilePath| option and use it as a working folder.
\begin{cautionbox}
When using the default setting, 
this option has to contain a process ID as well 
in order to work correctly 
when running \SLAM{}  in parallel on a computer cluster.
\end{cautionbox}
}
\item{\verb&Model&\index{Options!\verb&Model&}~\\
    selects either between different predefined scenarios indicated by a special string 
    or allows direct input, specifying the scenario via an internal SLHA format.
\begin{itemize}
 \item{ {\verb&"..."&}\\
 For predefined scenarios all possible choices of allowed strings can be found in Tab.~\ref{TAB:PreimplementedModels}.}
 \item{ {\sc Direct input}\\
A minimal example for the direct, or user-defined model input, 
is given in Section~\ref{CHAP:SelfDefinedScenarios}.  For a better 
understanding of the used format the user may be referred to
Section~\ref{CHAP:InternalRepresentationfortheSLHA}. The format
enables the user to add block comments or block information, or even
entry comments which will be written to the SLHA input file and
saved in the data base.
However, it is very important that all values should be given in
their proper format.  That means, if one wants to provide an integer
it needs to have the \verb|Head| \verb|Integer|.  
If one wants to provide a floating point number, 
the \verb|Head| of the number should return
\verb|Real|.  Even if the number happens to be an integer one should
append a dot to make its \verb|Head| a \verb|Real|.  This is
because \SLAM{}  has to convert the given input to proper {\tt
  FORTRAN} readable strings.
In case one already has a proper SLHA input file, 
it is possible to automatically generate the direct input using the {\tt Read\-LesHouches\-Spectrum\-File} function (see Section~\ref{CHAP:ReadLesHouchesSpectrumFile} on Page~\pageref{CHAP:ReadLesHouchesSpectrumFile} for more details ), 
in order to avoid time-consuming retyping. }
\end{itemize}
}
\item{\verb&RemoveTemporaryFiles&\index{Options!\verb&RemoveTemporaryFiles&}~\\
  can be \verb|True| or \verb|False|.  When set to \verb|True|
  \SLAM{}  deletes all temporary generated files automatically after
  reading them.  When set to \verb|False| they are kept.
\begin{cautionbox}
 When set to \verb|False| 
 one can easily get wrong results 
 because \SLAM{}  may just load an old file 
 in the case where no new file was generated by a spectrum generator.
 Thus this setting should only be used when one wants to have a look into the files.
\end{cautionbox} 
}
\item{\verb&Silent&\index{Options!\verb&Silent&}
\begin{itemize}
 \item \verb|"True"|\\
 Will stop \SLAM{}  to print processing messages.
 \item \verb|"False"|\\
 Processing messages get printed.
\end{itemize}
}
\item{\verb&SpectrumGenerator&\index{Options!\verb&SpectrumGenerator&}~\\
Specifies the spectrum generator that should be used.
Currently the following generators can be used, when installed and configured correctly:
\begin{itemize}
 \item \verb|"spheno"|,
 \item \verb|"softsusy"|,
 \item \verb|"suseflav"|,
 \item \verb|"suspect"|.
\end{itemize}
}
\end{itemize}
With the given options one can fully control the call of spectra from generators.
However, the options steering the communication with a data base are still missing.
They constitute the last class of options which will be introduced now.
In Tab.~\ref{TAB:DataBaseOptions} all options can be found in alphabetic order, 
including their default values
\begin{table}[t]
\small\centering
\begin{tabular}{|c|c|c|}
\hline
Option & Default Value \\
\hline
\verb|ClearDataBase| & \verb|False|\\
\verb|ExtendDataBase| & \verb|True|\\
\verb|KeepSQLConnection|& \verb|True|\\
\verb|LoadCompleteSpectra|&\verb|False|\\
\verb|MaxSQLConnectionAttempts|& \verb|10|\\
\verb|RefreshDataBase|& \verb|False|\\
\verb|SQLDataBase|& see \verb|SQLDataBaseValue| p\pageref{ITM:SQLDataBaseValue}\\
\verb|SQLDataBaseOptions|& see \verb|SQLDataBaseOptionsValue| p\pageref{ITM:SQLDataBaseOptionsValue}\\
\verb|SQLDataBaseType|& see \verb|SQLDataBaseTypeValue| p\pageref{ITM:SQLDataBaseTypeValue}\\
\verb|SQLVarCharLength|& see \verb|SQLVarCharLengthValue| p\pageref{ITM:SQLVarCharLengthValue}\\
\verb|TableDeclarationList|& see section~\ref{CHAP:DataBaseTableLayout}\\
\verb|TableName|& \verb|"SPECTRATABLE"|\\
\verb|UseDataBase|& \verb|True|\\
\hline
\end{tabular}
\caption{\label{TAB:DataBaseOptions}List of options used to control the data base operations of \SLAM.}
\end{table}
and are discussed in this order:
\begin{itemize}
 \item{\verb&ClearDataBase&\index{Options!\verb&ClearDataBase&}
 \begin{itemize}
 \item \verb|False|\\ Results in no additional action.
 \item \verb|Only|\\ Makes \SLAM{}   to only clear the current data
   base table and nothing else.
 \item \verb|True|\\ Does the same as \verb|Only| but after that normal processing is
   carried out.
 \end{itemize}
 \begin{cautionbox}
  Clearing a data base table 
  leads to a complete loss of all stored data.
  Use this option carefully!
 \end{cautionbox}

 }
 \item{\verb&ExtendDataBase&\index{Options!\verb&ExtendDataBase&}
  \begin{itemize}
 \item \verb|False|\\ New spectra are not saved in the data base.
 \item \verb|True|\\ New spectra are saved in the data base.
 \end{itemize}
 }
  \item{\verb&KeepSQLConnection&\index{Options!\verb&KeepSQLConnection&}
 \begin{itemize}
 \item \verb|False|\\ Makes \SLAM{}  disconnect from the SQL server after every transaction.
 \item \verb|True|\\ Makes \SLAM{}  to hold the connection to the SQL server throughout a \Mathematica{} session.
 \end{itemize}
 This setting may be useful when one has more than one
 \Mathematica{} kernel running \SLAM{}   at the same time (parallel)
 but using only one SQL server.  Depending on the SQL server settings,
 only a maximum amount of clients may be allowed to connect to the
 server.  That means holding the connection without need may
 potentially block another client from getting its data.  However,
 establishing and closing a connection after every interaction may
 increase the network traffic to an unacceptable amount.  }
 \item{\verb&LoadCompleteSpectra&\index{Options!\verb&LoadCompleteSpectra&}
 \begin{itemize}
 \item \verb|True|\\ Makes \SLAM{}  to load all data values contained
   in one spectrum and stored inside the data base table, although the
   user might only have requested some of the data values. However,
   only the requested data is returned after a filtering.
\item \verb|False|\\ Makes \SLAM{}  to load only the requested data
  values from the data base.  This requires some internal transformation
  of the {\tt Input\-Request} data to a SQL readable form, which might
  slow down the process for large requested sets of data.  However,
  with this setting the network traffic is reduced in any case and for
  a small number of requested data values this option leads to much
  faster loading times.
 \end{itemize}
 One should keep the default value (\verb|False|) in case of a small number of data values in a request.
 For requests close to a full data set the option \verb|True| might speed up the calculation
 because there is no processing time going into the conversion from {\tt Input\-Request} data to a SQL readable form.
 }
  \item{\verb&MaxSQLConnectionAttempts&\index{Options!\verb&MaxSQLConnectionAttempts&}~\\
  Can be any positive integer and sets the total amount of connection attempts to an SQL server, 
  until \SLAM{}   stops the operation throwing an error message.   
 }
  \item{\verb&RefreshDataBase&\index{Options!\verb&RefreshDataBase&}
 \begin{itemize}
 \item \verb|False|\\ Makes \SLAM{}  to load any matching spectra from the data base.
 \item \verb|True|\\ Makes \SLAM{}  to overwrite (refresh) matching spectra in the data base with the data of a newly generated spectrum.
 \end{itemize}
 }
  \item{\verb&SQLDataBase&\index{Options!\verb&SQLDataBase&}~\\
  Contains the address of the used data base as a string.
  In case the \Mathematica{} internal data base is used,
  this string is given by the system path where \Mathematica{} will write its data base files to.
  See \verb|SQLDataBaseValue| on Page~\pageref{ITM:SQLDataBaseValue} for examples.
 }
  \item{\verb&SQLDataBaseOptions&\index{Options!\verb&SQLDataBaseOptions&}~\\
  Contains a list of options which specify mainly the user account on the used data base.
  See {\tt SQL\-Data\-Base\-Options\-Value} on Page~\pageref{ITM:SQLDataBaseValue} for examples.
 }
  \item{\verb&SQLDataBaseType&\index{Options!\verb&SQLDataBaseType&}~\\
  Contains a string defining the kind of data base in use.
  See {\tt SQL\-Data\-Base\-Type\-Value} on Page~\pageref{ITM:SQLDataBaseTypeValue} for examples.
 }
  \item{\verb&SQLVarCharLength&\index{Options!\verb&SQLVarCharLength&}~\\
  Gives the maximum number of characters contained in a string saved to the data base as \verb|VARCHAR|.
  This option is only in use when a new data base is being created.
 }
  \item{\verb&TableDeclarationList&\index{Options!\verb&TableDeclarationList&}~\\
  Gives a list of declarations for the data base.
  The default value is generated automatically from the settings given in the file {\tt SLAM.config.m} when the \SLAM{}  package gets loaded.
  One should not use any option other than default, except for testing reasons.
  See Section~\ref{CHAP:DataBaseTableLayout} for more details.
 }
  \item{\verb&TableName&\index{Options!\verb&TableName&}~\\
  Holds a string which gives the name of the table inside the data base where all the spectra data are stored.
  Note that one might use different table names for different scenarios for convenience,
  especially for later manual analysis which may be independent of \SLAM.
  However, \SLAM{}  can distinguish different scenarios on its own, so saving spectra obtained within different scenarios does not lead to wrong results.
 }
  \item{\verb&UseDataBase&\index{Options!\verb&UseDataBase&}
 \begin{itemize}
 \item \verb|False|\\ Forbids \SLAM{}  to use the data base. Thus any spectrum request has to be served by a spectrum generator call.
 \item \verb|True|\\ Allows \SLAM{}  to use the data base. This means any matching spectrum found in the data base will be used, instead of calling a spectrum generator.
 \end{itemize}
 }
\end{itemize}

We have now described all relevant options,
we need to state some important usage restriction:

\noindent\fbox{\begin{smallercautionbox}
Although the SLHA allows for the printout of scale dependent parameters at different scales $Q$ in a single SLHA file,
\SLAM{}  is not able to treat multiple appearances of a block at different scales.
That means one should use the spectrum generators in user-defined scenarios in such a way, 
that each block is appearing only once per SLHA file.
\end{smallercautionbox}}

If one needs parameters at different scales, 
the spectra have to be called separately for each scale.

Since we have now all the basics at hand,
we can show some more involved applications of our package.

\subsection{Advanced Example}
A nice example for the comfortable usage
is the calculation of the light Higgs boson mass in a scenario depending on two parameters, 
like the $m_h^{\rm max}$ scenario.
How easy it is to get the relevant data shows the following code:

\MathematicaBox{\figpath MhMaxPlotData.pdf}

With the first command the default options of {\tt Obtain\-Les\-Houches\-Spectrum} get changed.
We select the $m_h^{\rm max}$-scenario, 
make the calculation silent, 
which means printouts are disabled, chose \softsusy{}  as the spectrum generator and request only the light Higgs boson mass to be in the output.
Changing the default values has the advantage, 
that we do not need to specify the options in the call of {\tt Obtain\-Les\-Houches\-Spectrum} itself, 
which keeps the next command clear and short.

In the second command, we generate the Higgs mass data in dependence of the two free parameters $M_A$ and $\tan\beta$.
This is done with the help of a simple \verb|Table| function, 
which automatically creates all possible combinations of $M_A$ and $\tan\beta$ values in the stated limits.
Note that for each combination {\tt Obtain\-Les\-Houches\-Spectrum} is called with the current value of the two parameters.
Since this function returns just a replacement rule for \verb|Mh0|, 
one has to use the called function like a replacement rule acting on the symbol \verb|Mh0|.
The next to outer-most \verb|Flatten| function reduces the structure to a list of lists, where each of the lists contains three elements:
the x-value $M_A$, the y-value $\tan\beta$ and the z-value $M_{h^0}$.
The outer-most {\tt Absolute\-Timing} function just returns the absolute calculation time in seconds of the command in its argument,
which was about 26 seconds, using the \Mathematica{} built-in data base.
This runtime was consumed to run the spectrum generator and fill the data base with data from $99$ spectra.

It is straightforward to display the calculated data, e.g. the following code does this job in a minimal way:

\MathematicaBox{\figpath MhMaxPlot.pdf}

The plot shows the Higgs mass in GeV in dependence of $M_A$ and $\tan\beta$.

In this example we started with an empty data base and the spectrum
generator had to run for every combination of input values. We can
rerun the command with the already populated data base 

\MathematicaBox{\figpath MhMaxPlotData2.pdf}

and find that the processing time now is about ten times smaller.

\subsection{Using {\tt Write\-LesHouches\-File}}
\index{Functions!{\tt Write\-LesHouches\-File}} The function {\tt
  Write\-LesHouches\-File} is normally called by {\tt
  Obtain\-Les\-Houches\-Spectrum}, but because it might be useful to be
able to write SLHA files directly from \Mathematica{} and independent
of the spectrum generator usage, this function can be called
independently.  The function has only three options, which shall be
discussed now:
\begin{itemize}
  \item{\verb&InputFilePath&\index{Options!\verb&InputFilePath&}~\\
  Holds a string which contains the path and the name of the file.
 }
  \item{\verb&InputList&\index{Options!\verb&InputList&}~\\
  Contains the actual data that gets written to the file.
  An example for the formatting of the data can be displayed with:
  \begin{verbatim}
  InputList/.Options[WriteLesHouchesFile]
  \end{verbatim}
  Further information about the format can be found in Section~\ref{CHAP:InternalRepresentationfortheSLHA}.
 }
 \item{\verb&AppendToFile&\index{Options!\verb&AppendToFile&}~
 \begin{itemize}
 \item \verb|False|\\ {\tt Write\-LesHouches\-File} clears already existing content in the target file before writing output to it.
 \item \verb|True|\\ {\tt Write\-LesHouches\-File} appends its output after already existing content in the target file.
 \end{itemize}
 }
\end{itemize}
With the given function one can write arbitrary data to a SLHA file, 
as the following example shows:

\MathematicaBox{\figpath WriteLesHouchesFile1.pdf}

This example leads to the output file:

\MathematicaBox{\figpath WriteLesHouchesFile2.pdf}

This shows that the application of {\tt Write\-LesHouches\-File} is not restricted to the spectrum generator interface.
Note that block information (\verb&BI&), block comments (\verb&BC&) and entry value comments (\verb&EC&) are optional in the \verb&InputList&.
Note further that {\tt Write\-LesHouches\-File} automatically detects the format of the given entry value (\verb&EV&).
All other values and comments have to be strings.

\subsection{Using {\tt Read\-LesHouches\-Spectrum\-File}}
\index{Functions!{\tt Read\-LesHouches\-Spectrum\-File}}
\label{CHAP:ReadLesHouchesSpectrumFile}
The function {\tt Read\-LesHouches\-Spectrum\-File}, like its name suggests, reads in SLHA files.
It has several options which are listed in alphabetic order including their default values in Tab.~\ref{TAB:ReadLesHouchesSpectrumFileOptions}.
\begin{table}[t]
\small\centering
\begin{tabular}{|c|c|c|}
\hline
Option & Default Value \\
\hline
\verb|AdoptEntryFormats| & \verb|True|\\
\verb|DataStructure| & \verb|LesHouchesOutputSQLFormats|$^{\dag}$\\
\verb|Input|& \verb|File|\\
\verb|InputRequest|& \verb|LesHouchesInputRequest|$^{\dag}$ \\
\verb|RemoveTemporaryFiles|& \verb|True|\\
\verb|OutputFilePath|& \verb|OutputFilePathValue|$^{\dag}$ \\
\hline
\end{tabular}
\caption{\label{TAB:ReadLesHouchesSpectrumFileOptions}List of all options of the function {\tt Read\-LesHouches\-Spectrum\-File}.
$^\dag$defined in {\tt SLAM.config.m}.}
\end{table}
The options and their descriptions are:
\begin{itemize}
  \item{\verb&AdoptEntryFormats&\index{Options!\verb&AdoptEntryFormats&}
 \begin{itemize}
 \item \verb|False|\\ Will keep the string format of all entry values read in.
 \item \verb|True|\\ Will convert the entry format according to either {\tt Input\-Request} or {\tt Data\-Structure}, depending on the value of {\tt Input\-Request}.
 \end{itemize}}
  \item{\verb&DataStructure&\index{Options!\verb&DataStructure&}~\\
  Gives the default declaration of all entry values.
  Its default value is given by \verb|LesHouchesOutputSQLFormats| in the {\tt SLAM.config.m} file.
  Note, that this can be changed to what is required for the content of the SLHA file getting loaded.
 }
 \item{\verb&Input&\index{Options!\verb&Input&}
 \begin{itemize}
 \item \verb|File|\\ Tells {\tt Read\-LesHouches\-Spectrum\-File} to read in the file according to the path and file name provided in \verb|OutputFilePath|.
 \item Direct string input \\ Will be read like stemming from a file.
 \end{itemize}
 }
\item{\verb&InputRequest&\index{Options!\verb&InputRequest&}
\begin{itemize}
\item{ \verb|All|\\
returns everything without touching the data format, that means everything stays a string.
}
\item{\verb|AllFormated|\\
Returns only blocks of entry values 
which are declared in the \verb&DataStructure& option.
Further, all values are converted to their declaration formats, if \verb&AdoptEntryFormats& is set to \verb|True|.
Use this option value in order to generate direct input useable as \verb&Model& option value of {\tt Obtain\-Les\-Houches\-Spec\-trum}. 
}
\item{{\sc Direct input}\\
Only the selected input values will be returned in a replacement rule,
working like for {\tt ObtainLesHouchesSpectrum} (See page~\ref{ITM:InputRequest}). 
}
\end{itemize}}
 \item{\verb&RemoveTemporaryFiles&\index{Options!\verb&RemoveTemporaryFiles&}
 \begin{itemize}
 \item \verb|True|\\ Tells {\tt Read\-LesHouches\-Spectrum\-File} to delete the file after reading it.
 \item \verb|False|\\ Will keep the file unchanged after reading it. 
 \end{itemize}
 }
\item{\verb&OutputFilePath&\index{Options!\verb&OutputFilePath&}~\\
Provides a string holding the path and the name of the file that should be read.
 }
\end{itemize}

Because we already have written out an example SLHA file in the previous section,
we can give an usage example of the function {\tt Read\-LesHouches\-Spectrum\-File} by just loading in, what we have written out:

\MathematicaBox{\figpath ReadLesHouchesSpectrumFile1.pdf}
\label{MB:ReadLesHouchesSpectrumFile1}

In the code above we set the default value of {\tt Remove\-Temporary\-Files} to \verb&False& to be sure that we can load in the input file again.
Further we specify the file which should be loaded.
After that we call the {\tt Read\-LesHouches\-Spectrum\-File} function, 
requesting everything that can be found in the SLHA file by setting the option {\tt Input\-Request -> All}.
We show the output in the {\tt Input\-Form} to be able to distinguish strings from regular symbol names.
The output shows everything that is contained in the file, but the entry values (\verb|EV|) are kept as strings.

In order to adopt their original format, 
we can load the file with the option {\tt Input\-Request -> AllFormated}.
Because the default option of {\tt Data\-Structure}, 
which is given by {\tt Les\-Houches\-Output\-SQL\-Formats},
does not contain any suitable declaration information for our little toy example,
we have to provide the proper declaration through the {\tt Data\-Structure} option on our own.
This is done in the following:

\MathematicaBox{\figpath ReadLesHouchesSpectrumFile2.pdf}
\label{MB:ReadLesHouchesSpectrumFile2}

Compared to the output of Mbx.~\refm{MB:ReadLesHouchesSpectrumFile1}, 
we see that all entry values 
which have been declared 
got converted from string to the declared format.
Since we did not declare anything for the block \verb&"HOLIDAYS"&, 
this block is completely missing.

In the last example we directly request just the number of advents in block \verb&"XMAS"&:

\MathematicaBox{\figpath ReadLesHouchesSpectrumFile3.pdf}
\label{MB:ReadLesHouchesSpectrumFile3}

This results in a list containing a replacement rule, 
where our choice of symbol for the number of advents has been taken over and the number was properly converted to an integer.

\section{Package Internals}
in this section we provide information, which enables the user to extend or adjust \SLAM{} to very special needs.
Before going into details, some words about the general layout of the package and the spirit in which it has been programmed are in order.

The package is purely written in \Mathematica.
Only very little \bash{}  code is used to manage the spectrum generators and the SLHA files.
Throughout the package the code was written in a modular way, 
heavily relying on the built-in options system of \Mathematica. A
comprehensive introduction to the options interface can be found in
the \Mathematica{} help centre.
It is easy to write argument insensitive code for
the call of a sub function inside a normal function, 
if one uses the options interface.  One does not need
to worry about which argument has to be put at which position and it
is no problem to add further parameters later, without changing the
code for the function call itself.  With this feature one can
conveniently write modular code, grouping different tasks into
different, topic oriented functions.  This helps to keep the code
local and modular, which is very convenient while debugging.  If
something is not working properly, one can just check the data exchanged
between modules.  Finding the place where wrong data appears leads
directly to the faulty module.  Moreover, since every module serves a
very special purpose only, it is simpler to ensure that each module
fulfils its task properly than just writing a single code which has to
achieve many goals properly at the same time.  Thus following the old
maxim ``divide et impera''\footnote{lat. for ``divide and rule''}, one
can write complicated code achieving multiple goals properly by just
writing small modules doing their job properly and finally connecting
all of them.

In the following subsection we are going to give a ``map'' of all relevant main modules
which work inside the {\tt Obtain\-LesHouches\-Spectrum} function.
We will walk on this map through the steps which are performed automatically by \Mathematica{} in order to carry out commands given by the user.
During this subsection the reader should get a basic orientation, needed for any modification of the code.

The second subsection clarifies the \Mathematica{} internal representation of the SLHA used in this package. 

In the third subsection we explain the used layout of the data base table in more detail.

The fourth subsection should enable the reader to implement new predefined scenarios in the existing code of \SLAM. 

\subsection{Internal Structure of {\tt Obtain\-LesHouches\-Spectrum}}
\begin{figure}[t]
\includegraphics[scale=0.5]{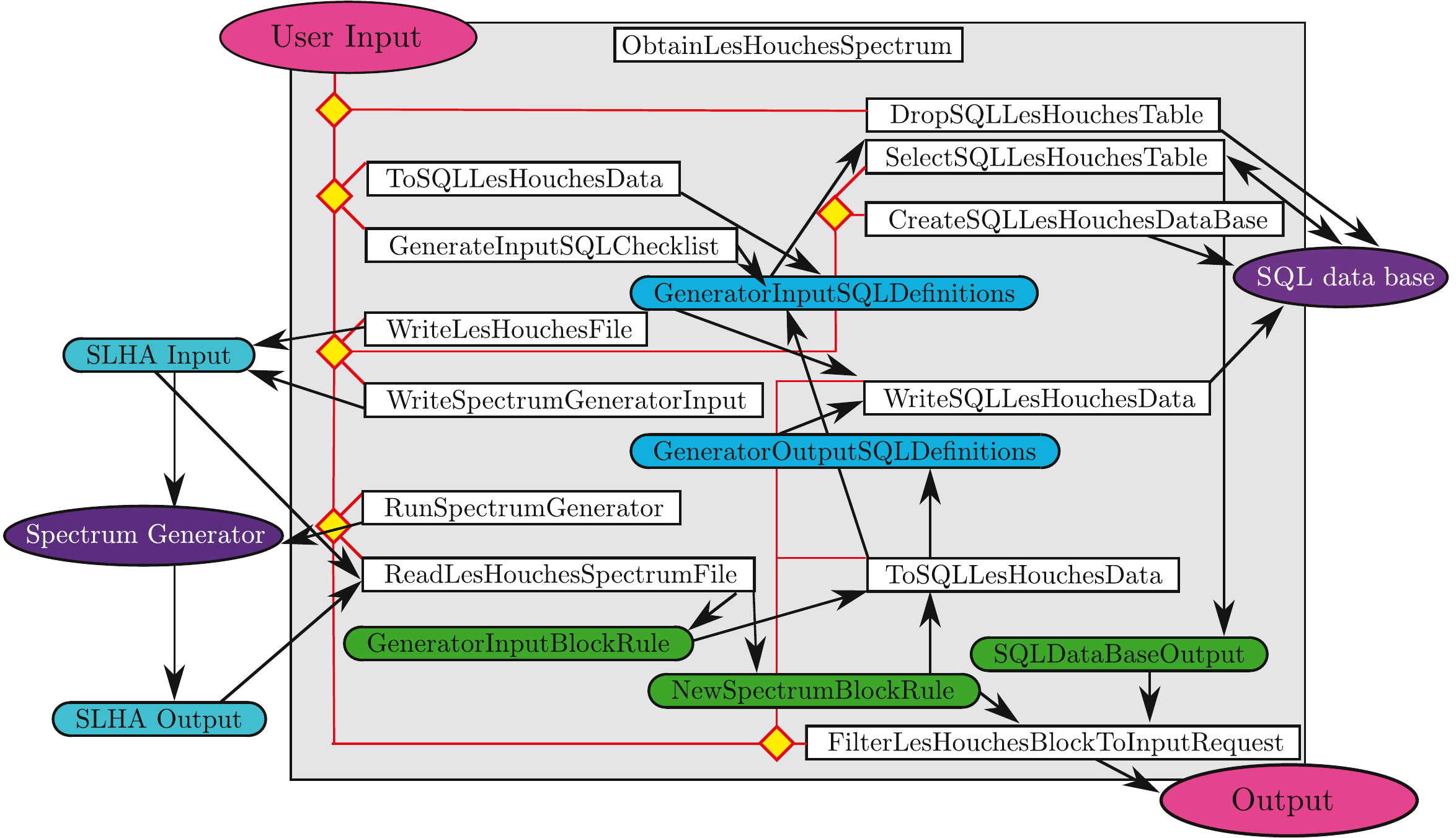}
\caption{\label{FIG:ObtainLesHouchesSpectrum}Internal structure of the
  function {\tt Obtain\-LesHouches\-Spectrum}.  The named white
  rectangular boxes correspond to sub functions used by {\tt
    Obtain\-LesHouches\-Spectrum}.  Arrows indicate the data flow.
  Rounded boxes stand for data in certain format, depending on their
  colour.  The red lines follow the order of function calls.  Yellow
  diamonds show possible branchings into different function calls in
  dependence of certain conditions.}
\end{figure}
In Fig.~\ref{FIG:ObtainLesHouchesSpectrum} a simplified ``blueprint''
of the function {\tt Obtain\-LesHouches\-Spectrum} is shown.  The
function connects a spectrum generator on the left and a SQL data base on
the right with the output of the function at the bottom right, in
dependence of input provided by the user, shown at the top left.
Inside the function {\tt Obtain\-LesHouches\-Spectrum}, which is
indicated by the light grey background, sub functions are displayed in
white rectangular boxes.  They are connected by red lines indicating
the order of sub-function calls.  The yellow diamonds represent
branchings of function calls in dependence of certain conditions.  The
rounded boxes stand for data prepared in a certain format indicated by
the colour of the background.  In fact all names inside are actual
function and data names used in the \Mathematica{} code.  Black arrows
show the flow of data.

In the following the flowchart is described step by step. 
The starting point of this description is the place where the user input enters {\tt Obtain\-LesHouches\-Spectrum}:
\begin{enumerate}
 \item In dependence of the option \verb|ClearDataBase| the SQL data base table will be removed by the sub function {\tt Drop\-SQL\-LesHouches\-Table} when \verb|True| or \verb|Only|.
 \item In case of \verb|Only| the evaluation ends after that. \XBox
 \item When \verb|True| or \verb|False| was selected it generates the data {\tt Generator\-Input\-SQL\-Definitions} with either the {\tt To\-SQL\-LesHouches\-Data} or the {\tt Generate\-Input\-SQL\-Checklist} function.
       The first one is used in case a direct input model is used, the second function will do the job when a predefined model is considered.
 \item{ When the option {\tt Use\-Data\-Base} is \verb|True|, the next step is to check if there is already any data consistent with the defined {\tt Generator\-Input\-SQL\-Definitions} in the data base.
       This is done by calling the sub function {\tt Select\-SQL\-LesHouches\-Table}.
       \begin{enumerate}
       \item If there is any spectrum ready, it is stored in {\tt SQL\-Data\-Base\-Output}
             which gets filtered by the sub function {\tt Filter\-LesHouches\-Block\-To\-Input\-Request} in order to provide the final output and end the evaluation. \XBox
       \item In case no proper data base table could be found it is created by the sub function {\tt Create\-SQL\-LesHouches\-Data\-Base}. Further steps coincide with the case below.
       \item If no match was found, the spectrum needs to be obtained from the spectrum generator.
       \end{enumerate}
       }
 \item{In order to obtain a spectrum from the generator, the proper SLHA input file has to be created.
 This is either done with the sub function {\tt Write\-LesHouches\-File}, 
 when a direct input model is considered, 
 or with the sub function {\tt Write\-Spectrum\-Generator\-Input}, 
 when a predefined model is in use.}
 \item{Then the spectrum generator is run by the sub function {\tt Run\-Spectrum\-Generator}.}
 \item{If there was no problem in the previous step, the sub function {\tt Read\-LesHouches\-Spectrum\-File} is used to read in the SLHA output file and store the data in the expression {\tt New\-Spectrum\-Block\-Rule}.}
 \item{In case the options {\tt Extend\-Data\-Base} and {\tt Use\-Data\-Base} were set to \verb|True|:
 \begin{enumerate}[8.1]
  \item {\tt Read\-LesHouches\-Spectrum\-File} is used to read in the SLHA input file and store the data in the expressions {\tt Generator\-Input\-Block\-Rule}.
  \item {\tt To\-SQL\-LesHouches\-Data} transforms the data in {\tt Gen\-er\-at\-or\-Input\-Block\-Rule} and {\tt New\-Spectrum\-Block\-Rule} to the data base writable expressions {\tt Generator\-Input\-SQL\-Definitions} and {\tt Generator\-Output\-SQL\-Definitions}. 
  \item Then both sets of data are saved together in the \SQL{}  data base through the sub function {\tt Write\-SQL\-LesHouches\-Data}. 
 \end{enumerate}
  }
  \item{The final step is done by the sub function {\tt Filter\-LesHouches\-Block\-To\-Input\-Request} filtering the requested output from the data of {\tt New\-Spectrum\-Block\-Rule} in order to return it as output. \XBox}
\end{enumerate}
We did not discuss any option relevant for the sub functions.
For example, the {\tt Spectrum\-Generator} option, 
telling {\tt Obtain\-LesHouches\-Spectrum} which spectrum generator should be used, 
does not appear in the code of the function at all.
This is because {\tt Obtain\-LesHouches\-Spectrum} just trades these options to its sub function via a generic interface.
This keeps the structure of the code very simple and generic.

\subsection{Internal Representation for the SLHA}\label{CHAP:InternalRepresentationfortheSLHA}
In the following subsection we explain the \Mathematica{} internal format of SLHA data, 
which is in use for the expressions {\tt Generator\-Input\-Block\-Rule}, {\tt New\-Spectrum\-Block\-Rule} and {\tt SQL\-Data\-Base\-Output} (see Fig.~\ref{FIG:ObtainLesHouchesSpectrum}).
The easiest way of explaining a certain format is to give just a data example.
A SLHA output like:
\begin{verbatim}
Block MINPAR  # Input parameters
 3   2.00000000E+01  # tanb at m_Z
 4   1.00000000E+00  # Sign(mu)
Block ALPHA  # Effective Higgs mixing angle
    -1.05379330E-01  # alpha
Block YU Q=  1.01490536E+03  # (SUSY scale)
 1 1  8.59529004E-06  # Y_u(Q)^DRbar
\end{verbatim}
looks like the following in the \Mathematica{} internal notation:
\begin{verbatim}
 ExampleSLHAData={
 "MINPAR" -> {
      BC -> " Input parameters", 
      BI -> {EV -> None, EC -> None}, 
      3 -> {EV -> "2.00000000E+01", EC -> "tanb at m_Z"}, 
      4 -> {EV -> "1.00000000E+00", EC -> "Sign(mu)"}}, 
 "ALPHA" -> {
      BC -> " Effective Higgs mixing angle", 
      BI -> {EV -> None, EC -> None}, 
      None -> {EV -> "-1.05379330E-01", EC -> "alpha"}},
 "YU" -> {
      BC -> " (SUSY scale)", 
      BI -> {EV -> "Q=  1.01490536E+03 ", EC -> None}, 
      {1, 1} -> {EV -> "8.59529004E-06", EC -> "Y_u(Q)^DRbar"}}};
\end{verbatim}
As one can see, the SLHA data is ordered with respect to blocks, 
where each block is just a replacement rule giving the block name in capital letters as string on the left hand side of the rule 
and the content of the block in a list on the right hand side.

Each of those lists contains a replacement rule for the block comment (\verb|BC|).
Further it contains a replacement rule for the block information (\verb|BI|), which has a list of two replacements on its right hand side.
The latter contains a replacement for its entry value (\verb|EV|) and entry comment (\verb|EC|), where in case of the block information the last one is always \verb|None|.

In case the entry value is given by a matrix element depending on two entry values, instead of a single integer as entry key, a list of two integers is used,
as can be seen in the last line.

In case where there is no entry key for a value, like for the mixing angle $\alpha$, 
the variable \verb|None| is used as entry key.

Note that in the example above all entry values are kept as strings.
This only happens when SLHA files are loaded from a generator 
using the option {\tt InputRequest->All}.

In case of {\tt InputRequest->AllFormated}~\SLAM{}  tries to return converted entry values (e.g. real numbers) 
using the declarations made in the file {\tt SLAM.config.m}. 
\begin{cautionbox}
However, blocks which were not declared at all will not be displayed in the output.
\end{cautionbox}

Entry values are returned in their declared format when loading spectra from the data base. 

\subsection{Data Base Table Layout}\label{CHAP:DataBaseTableLayout}
Because one can use the automatically created and built-up data base independently from \SLAM,
the layout of the table should be clarified.
The default option value for \verb|TableName| leads to the creation of a table with name \verb|"SPECTRATABLE"|.

Any table is declared and initialized with the information provided in the variables {\tt LesHouches\-Input\-SQL\-Formats} and {\tt LesHouches\-Output\-SQLFormats} defined in the file {\tt SLAM.config.m}.
These variables are read in and the contained information is processed once, 
when the \SLAM{}  Package gets loaded, to give the full declaration list of the table.
The full list of declarations can be printed  using the command:
{\small\begin{verbatim}
 (TableDeclarationList /. Options[ObtainLesHouchesSpectrum]) // TableForm
\end{verbatim}}
We do not give the full default output here because it is quite lengthy, but pick only a selection for demonstration:

 \MathematicaBox{\figpath tabledeclarations.pdf}
 \label{MB:tabledeclarations}
 
Note that all terms appearing in this output are in fact strings, 
because \verb|TableForm| does not display the \verb|"| character.

The output shows pairs of strings 
where the first string gives the name of the corresponding column and encodes mainly the place where the data is located in the SLHA file.
The second string defines the type of data 
which will be stored in the corresponding column.
So far there are only three different data types in use: \verb|"INTEGER"|, \verb|"DOUBLE"| and \verb|"VARCHAR"|.

In the selection shown in Mbx.~\refm{MB:tabledeclarations} one can distinguish two different kinds of column names:
\begin{itemize}
 \item Column names starting with a ``\verb|I|'' reserve space for data stemming from SLHA input files.
 \item Column names starting with a ``\verb|O|'' reserve space for data generated by the spectrum generators, so stemming from the SLHA output files.
\end{itemize}
Further the column names ending with \verb|COMMENT| lead to columns, 
which do only save comments in the \verb|"VARCHAR"| format.
This is because one can optionally place comments behind the hash symbol (\verb|#|) in a SLHA file in each line.

After the initial ``\verb|I|'' or ``\verb|O|'' comes a block name in capital letters separated by an underscore.

\begin{itemize}
 \item{ 
For columns which do not save a string one of the following holds then:
\renewcommand{\theenumi}{\alph{enumi})}
\renewcommand{\labelenumi}{\theenumi}
\begin{enumerate}
 \item At least one further underscore separated integer value is required.
       This value corresponds to the key value defined in the SLHA.
       See lines seven and nine in the displayed output above for example.
 \item If the value represents a matrix element 
       there may be two underscore separated integers.
       Those integers correspond to two valued key entries defined in the SLHA.
 \item If the corresponding SLHA quantity has no key value, 
       that means it comes alone in its own data block, 
       like it is the case for the mixing angle $\alpha$,
       the string \verb|NONE| has to follow the block name after one underscore.  
\end{enumerate}
\renewcommand{\theenumi}{\arabic{enumi}.}
\renewcommand{\labelenumi}{\theenumi}
}

\item{Every data value entry can have its own comment.
The name of the comment column is obtained by simply appending the string \verb|_COMMENT| to the name of the corresponding data value column.
See lines ten and eight in the output above for example.
}
\item{Every block can have its own comment which is saved in the column name given by the block location appending the string \verb|_COMMENT|.
Lines one, three and five of the displayed output above give examples for this case.}
\item{Every block can have an additional information entry 
which is mainly used to hold the scale information.
For example a block definition in a SLHA file might look like the following:
\begin{verbatim}
 BLOCK GAUGE Q=  1.01490536E+03  # (SUSY scale)
\end{verbatim}
Everything after the space behind \verb|GAUGE| and before the hash symbol (\verb|#|) gets stored in the block information (\verb|BI|).
The information is automatically saved in the \verb|"VARCHAR"| format, 
as can be seen e.g. from lines two, four and six in the printout above
where the name of the column is ending with the string \verb|_INFO|.
Note that requesting this data in the output as \verb|"Real"| will force
\SLAM{}  to return just the real number without the \verb|Q=|.
}
\end{itemize}
With the given information it is easy to search directly in the data base for certain parameter limits.
An example is to search for Higgs boson masses between $123$ GeV and $129$ GeV:
\begin{verbatim}
select I_EXTPAR_26 as MA,
       I_MINPAR_3 as tanbeta,
       O_MASS_25 as MH  
       from SPECTRATABLE 
       where (O_MASS_25 > 123 && O_MASS_25 < 129) 
       order by MH;
\end{verbatim}

{\tt }

\subsection{Creating Additional Predefined Scenarios}
Adding a new predefined scenario requires 6 steps but may help to speed up the generation of spectra, 
because less calculation steps have to be done for predefined scenarios in \Mathematica.
\begin{enumerate}
 \item If the new  predefined scenario requires new option values 
 because it depends on parameters which have not been used yet,
one should add a new usage instruction before the {\tt Private} section in the file {\tt SLAM.m} for every symbol which will become an option. 
For example, the usage instruction for the \verb|nmess| parameter used in the mGMSB-scenario looks like:
\begin{verbatim}
nmess::usage="Number of minimal copies of the messenger 
sector in the gauge mediated symmetry breaking model."; 
\end{verbatim}
This enables the user to get information for that option via a question mark operation like:
\verb|?nmess|.
\item Add the option name and description of the new scenario to the usage instruction of the {\tt Model} variable:
\begin{verbatim}
Model::usage="..."
\end{verbatim}
For example the additional line should look like:
\begin{verbatim}
\"mgmsb\" (gauge mediated symmetry breaking).
\end{verbatim}
\item Add the definitions for the default option value symbols related to the new option value in the {\tt SLAM.config.m} file. 
The names of the default option value symbols are just obtained from the option name itself by appending {\tt Value}.
See for example the \verb|nmessValue| in Tab.~\ref{TAB:DefaultValues}.
That way, the file {\tt SLAM.m} will stay clean of any default numerical value definition and any user has a nice collection of the default numerical values in the config file. 
\item Add the new options to the options of {\tt Write\-Spectrum\-Generator\-Input}.
For our \verb|nmess| example the corresponding line, which would have to be added, looks like:
\begin{verbatim}
 nmess -> Global`nmessValue,
\end{verbatim}
\item The most difficult part is to add the new scenario consistently
  to the subfunction {\tt Write\-Spectrum\-Generator\-Input}.  There
  are multiple \verb|Switch| instructions where the new code for the new
  scenario has to be added.  This can be done by just following the
  given examples in the code itself.  Once having finished the
  modification of this function, one should check, after reloading the
  package, that the string written to the input SLHA file is correct
  when calling the new scenario.  An easy way to do that is to call
  the modified function directly via:
\begin{verbatim}
SLAM`Private`WriteSpectrumGeneratorInput[Model -> "mgmsb",
InputFilePath -> String]
\end{verbatim}
where one of course has to replace \verb|"mgmsb"| by the option value
of the new scenario.
\item In the final step the new  scenario has to be added to the body of the sub function {\tt Generate\-Input\-SQL\-Checklist}.
Note that this works pretty much the same as in the previous step.

In fact all numerical values entered directly {\it{have}} to be equal to those entered in the previous step!
If this is not the case \SLAM{}  may not find anything in the data base although it already saved a requested spectrum.
Please note that one should apply the function \verb~StableNumber~ to real numbers 
in order to make them stable under the conversion from and to their \verb~FORTRAN~ form.
If a number is not stable under these conversions,
very small rounding errors occur,
which spoil the detection of matching spectra in the data base.

Each numerical value follows an identification string in a list.  This
string can be built up from the position of the numerical value in a
SLHA file as follows: Conventionally the first letter is a
``\verb|I|'' which indicates that the value is entering the input
value part of the \SQL{} data base.  Since white spaces are not
allowed in \SQL{} to be part of a column name, the separation to the
following block name in capital letters is done by an underscore.
After the block name the underscore separated key value follows as an
integer.  If it is a double key value, like used for matrix elements,
there may be two integer numbers separated by one underscore.  If
there is no entry key at all, like it is the case for the output value
of the mixing angle $\alpha$, which somehow got its own block,
\verb|NONE| replaces the usual integer number.  One can check the
adjustments by directly calling the function via:
\begin{verbatim}
 SLAM`Private`GenerateInputSQLChecklist[Model -> "mgmsb"]
\end{verbatim}
where \verb|"mgmsb"| has to be replaced by the option value of the new scenario.
The output should be just a list of pairs, where each pair is a list with an identification string as the first element and a numerical value as the second element.
\end{enumerate}
Once the given steps have been completed, the new predefined scenario
should work without any problems.

\section{Summary}
We presented and published the package \SLAM{},
which provides a convenient interface for SLHA spectrum generators in \Mathematica{}.

The package enables the user to obtain spectrum data from generators
in a fully automatic way.  Results of different spectrum generators
can be compared without any effort and it allows the user to use 
his own notation in \Mathematica{}.  \SLAM{} comes with a
large number of built-in benchmark scenarios.  Furthermore, it allows the
user to freely define any desired scenario following the SLHA
standard.

Moreover, it can store and recall all acquired data to and from a data base in order to avoid a recalculation of known spectra.
Storing spectra in a data base allows the examination of parameter spaces 
by simply using powerful data base functionalities.
A parallel use of \SLAM{} is possible and helps to reduce the time needed for possible parameter scans and builds of data bases.

Besides the pure usage documentation including examples, 
we provided more details about the internal structure of the package 
which may help in case a modification of the program code
is needed due to special user requirements.



\section*{Acknowledgements}
We would like to thank M.~Iskrzynski and A.~Kurz for beta testing the application,  and J.~Hoff, A.~Kurz and M.~Steinhauser for reading the manuscript. 
Moreover we appreciate the fruitful time at the Institut f\"ur Theoretische Teilchenphysik at KIT,
where the core routines of \SLAM{} have been implemented and runtime benchmarks could be performed on the local PC cluster.
This work has been supported in part by the EU Network
{\sf LHCPHENOnet} PITN-GA-2010-264564, and by  DFG Sonderforschungsbereich Transregio 9,
Computergest\"utzte Theoretische Teilchenphysik.




\printindex


\begin{thebibliography}{99}

%
%

\bibitem{Porod:2003um}
  W.~Porod,
  Comput.\ Phys.\ Commun.\  {\bf 153} (2003) 275
  [hep-ph/0301101].

\bibitem{Allanach:2001kg}
  B.~C.~Allanach,
  Comput.\ Phys.\ Commun.\  {\bf 143} (2002) 305
  [arXiv:hep-ph/0104145].

\bibitem{Chowdhury:2011zr}
  D.~Chowdhury, R.~Garani and S.~KVempati,
  Comput.\ Phys.\ Commun.\  {\bf 184} (2013) 899
  [arXiv:1109.3551 [hep-ph]].

\bibitem{Djouadi:2002ze}
  A.~Djouadi, J.~-L.~Kneur and G.~Moultaka,
  Comput.\ Phys.\ Commun.\  {\bf 176} (2007) 426
  [hep-ph/0211331].

\bibitem{Skands:2003cj}
  P.~Z.~Skands, B.~C.~Allanach, H.~Baer, C.~Balazs, G.~Belanger, F.~Boudjema,
  A.~Djouadi and R.~Godbole {\it et al.},
  JHEP {\bf 0407} (2004) 036
  [hep-ph/0311123].

\bibitem{Allanach:2008qq}
  B.~C.~Allanach, C.~Balazs, G.~Belanger, M.~Bernhardt, F.~Boudjema,
  D.~Choudhury, K.~Desch and U.~Ellwanger {\it et al.},
  Comput.\ Phys.\ Commun.\  {\bf 180} (2009) 8
  [arXiv:0801.0045 [hep-ph]].

\bibitem{SLHAWeb}
  http://home.fnal.gov/$\sim$skands/slha/

\bibitem{1c++Web}
  http://fthomas.github.com/slhaea/

\bibitem{2c++Web}
  http://www.svenkreiss.com/SLHAio

\bibitem{LHPC}
  https://lhpc.hepforge.org/

\bibitem{Belanger:2010st}
  G.~Belanger, N.~D.~Christensen, A.~Pukhov and A.~Semenov,
  Comput.\ Phys.\ Commun.\  {\bf 182} (2011) 763
  [arXiv:1008.0181 [hep-ph]].

\bibitem{Hahn:2004bc}
  T.~Hahn,
  hep-ph/0408283.

\bibitem{Hahn:2006nq}
  T.~Hahn,
  Comput.\ Phys.\ Commun.\  {\bf 180} (2009) 1681
  [hep-ph/0605049].

\bibitem{pyslhaWeb}
  http://www.insectnation.org/projects/pyslha

\bibitem{Kant:2010tf}
  P.~Kant, R.~V.~Harlander, L.~Mihaila and M.~Steinhauser,
  JHEP {\bf 1008} (2010) 104
  [arXiv:1005.5709 [hep-ph]].

\bibitem{Pak:2012xr}
  A.~Pak, M.~Steinhauser and N.~Zerf,
  arXiv:1208.1588 [hep-ph].

\bibitem{Kurz:2012ff}
  A.~Kurz, M.~Steinhauser and N.~Zerf,
  arXiv:1206.6675 [hep-ph].

\bibitem{Carena:2002qg}
  M.~S.~Carena, S.~Heinemeyer, C.~E.~M.~Wagner and G.~Weiglein,
  Eur.\ Phys.\ J.\ C {\bf 26} (2003) 601
  [hep-ph/0202167].

\bibitem{AbdusSalam:2011fc}
  S.~S.~AbdusSalam, B.~C.~Allanach, H.~K.~Dreiner, J.~Ellis, U.~Ellwanger, J.~Gunion, S.~Heinemeyer and M.~Kraemer {\it et al.},
  Eur.\ Phys.\ J.\ C {\bf 71} (2011) 1835
  [arXiv:1109.3859 [hep-ph]].

\bibitem{SLAM_DL_LINK}
 http://www.ttp.kit.edu/Progdata/ttp13/ttp13-024/

\bibitem{Carena:2013qia}
  M.~Carena, S.~Heinemeyer, O.~Stål, C.~E.~M.~Wagner and G.~Weiglein,
  arXiv:1302.7033 [hep-ph].

\bibitem{Cahill-Rowley:2013gca}
  M.~W.~Cahill-Rowley, J.~L.~Hewett, A.~Ismail, M.~E.~Peskin and T.~G.~Rizzo,
  arXiv:1305.2419 [hep-ph].  


\end{thebibliography}
\end{document}